\documentclass[Physsubmission, Phys]{SciPost}

\hypersetup{
    colorlinks,
    linkcolor={red!50!black},
    citecolor={blue!50!black},
    urlcolor={blue!80!black}
}

\usepackage[bitstream-charter]{mathdesign}
\DeclareSymbolFont{usualmathcal}{OMS}{cmsy}{m}{n}
\DeclareSymbolFontAlphabet{\mathcal}{usualmathcal}

\graphicspath{{./eps/}}
\begin{document}

\begin{center}{\Large \textbf{T-violating effect in $\nu_{\tau} (\bar{\nu}_{\tau})-$nucleon quasielastic 
scattering
}}\end{center}

\begin{center}
A. Fatima\textsuperscript{1$\star$},
M. Sajjad Athar\textsuperscript{1} and
S. K. Singh\textsuperscript{1}
\end{center}

\begin{center}
{\bf 1} Department of Physics, Aligarh Muslim University, Aligarh-202002, India.\\
* atikafatima1706@gmail.com
\end{center}

\begin{center}
\today
\end{center}


\definecolor{palegray}{gray}{0.95}
\begin{center}
\colorbox{palegray}{
  \begin{minipage}{0.95\textwidth}
    \begin{center}
    {\it  16th International Workshop on Tau Lepton Physics (TAU2021),}\\
    {\it September 27 – October 1, 2021} \\
    \doi{10.21468/SciPostPhysProc.?}\\
    \end{center}
  \end{minipage}
}
\end{center}

\section*{Abstract}
{\bf The production cross sections and polarization observables of the $\tau$ leptons produced in the $|\Delta S| = 0$ and 
$1$ induced $\nu_{\tau}(\bar{\nu}_{\tau})-N$ quasielastic scattering have been studied. The effect of T violation, in the case 
of $\Delta S=0$ and 1 processes, and the SU(3) symmetry breaking effects, in the case of $\Delta S=1$ processes, on the total 
scattering cross sections as well polarization observables are explored. Experimentally, it would be possible to observe 
these effects in the forthcoming (anti)neutrino experiments like DUNE, SHiP and DsTau.
}


\section{Introduction}
\label{sec:intro}
In this work, we present a theoretical study of the production cross section as well as the polarization observables of the 
$\tau$ lepton produced in the quasielastic $\nu_{\tau} (\bar{\nu}_{\tau})-N$ scattering in the few GeV energy region relevant 
to the Deep Underground Neutrino Experiment~(DUNE), Search for Hidden Particles~(SHiP) and Study of tau neutrino production 
at the CERN-SPS~(DsTau) experiments. The $\tau$ lepton produced in $\nu_{\tau}-N$ scattering decays to leptons and pions 
through the leptonic and hadronic decay modes. In this energy region, the production cross section of $\tau$, its decay and 
the characteristics of the decay products depend significantly on the $\tau$ polarization. The production cross sections and 
polarization of $\tau$ leptons are calculated using various weak nucleon form factors which are determined using symmetry 
properties of the weak currents in the vector and axial vector sectors, assuming G and T invariances and SU(3) symmetry. The 
longitudinal and perpendicular components of polarization lie in the plane while the transverse component of polarization 
lies perpendicular to the reaction plane and is forbidden by G- and T-invariance. 
We have earlier studied the effects of G and T violation on the total cross section as well as on the polarization observables 
of the final nucleon/hyperon and the lepton produced in the quasielastic scattering of (anti)neutrinos with nucleons for 
both $\Delta S=0$ and $1$ processes~\cite{Fatima:2018tzs, Fatima:2020pvv, Fatima:2021ctt}. 
In the case of $\Delta S=1$ reactions, we study the effect of SU(3) symmetry breaking 
on these observables.

In Section~\ref{QE}, we discuss in brief the formalism for calculating the differential as well as total scattering cross 
sections for quasielastic scattering of $\nu_{\tau}(\bar{\nu}_{\tau})$ with nucleons. The polarization components of the 
produced $\tau$ lepton are discussed in Section~\ref{pol}. The SU(3) symmetry breaking effects, following the works of 
Faessler {\it et al.}~\cite{Faessler:2008ix} and Schlumpf~\cite{Schlumpf}, are discussed in Section~\ref{SU}. In 
Section~\ref{results}, we present and discuss the results for the total cross sections and average polarizations and 
Section~\ref{conclusion} concludes the present work.

\section{Quasielastic production of nucleons and hyperons}\label{QE}
The $\nu_{\tau}(\bar{\nu}_{\tau})$ induced quasielastic production on the free 
nucleon target are given by the reactions
\begin{eqnarray}\label{hyp-rec}
{\nu_{\tau}} (\bar{\nu}_{\tau})(k) + N(p)&\longrightarrow& \tau^{\mp}(k^\prime) + N(p^\prime),  ~~~~~N=n,p  \\
\bar{\nu}_{\tau}(k) + N(p)&\longrightarrow& \tau^{+}(k^\prime) + Y(p^\prime),  ~~~~~Y=\Lambda,\Sigma^0, 
\Sigma^{-}, 
\end{eqnarray} 
for which the transition matrix element is given by
\begin{equation}\label{matrix1}
{\cal{M}}=\frac{G_F}{\sqrt{2}} ~a~ l^{\mu}~J_\mu, 
\end{equation}
where $a = \cos \theta_{c}~(\sin \theta_{c})$ for $\Delta S=0~(1)$ processes and the 
leptonic ($l^{\mu}$) and the hadronic ($J_\mu$) currents are defined as
\begin{eqnarray}\label{lmu}
l^\mu &=& \bar u(k^\prime) \gamma^\mu (1 \mp \gamma_5) u(k), \\
 J_{\mu} &=& \bar{u}(p^{\prime})\left[\gamma_\mu f_1(Q^2)+i\sigma_{\mu\nu} \frac{q^\nu}{M+M_Y} f_2(Q^2)+ 
\frac{2 q_\mu }{M + M_Y} f_3(Q^2)\right.\nonumber  \\ 
\label{jmu}
&-& \left. \gamma_\mu \gamma_5 g_1(Q^2)- i \sigma_{\mu\nu}\gamma_5 
 \frac{q^\nu}{M+M_Y} g_2(Q^2) - \frac{2 q_\mu \gamma_5} {M + M_Y} g_3(Q^2)  \right] u(p).~~~~~
\end{eqnarray}
In Eq.~(\ref{lmu}), $+(-)$ stands for $\bar{\nu}_{\tau}(\nu_{\tau})$ induced process. In Eq.~(\ref{jmu}), 
$f_1 (Q^2)$, $f_2 (Q^2)$, $g_1 (Q^2)$ and $g_3 (Q^2)$ are the form factors associated with the first class currents while 
$f_3 (Q^2)$ and $g_2 (Q^2)$ are the form factors associated with the
second class currents. These form factors are determined using various symmetry properties of the weak hadronic current like 
G invariance, T invariance, SU(3) symmetry, etc. 
The real value of $g_2(0)$ gives G violation while T is conserved whereas the imaginary value of $g_2(0)$ gives G violation 
as well as T violation.

\subsection{Cross sections}
The $Q^2$ distribution is written as
\begin{equation}\label{dsig}
\frac{d\sigma}{dQ^2}=\frac{G_F^2 ~a^2}{8 \pi M^2 {E^2_{{\nu}_\tau(\bar{\nu}_{\tau})}}} N(Q^{2}),
\end{equation}
where the expression of $N(Q^{2})$ is given in Ref.~\cite{Fatima:2018tzs}. The expression for the total cross section is 
obtained by integrating Eq.~(\ref{dsig}) over $Q^2$ as:
\begin{equation}\label{sig}
\sigma= \int \frac{G_F^2 ~a^2}{8 \pi M^2 {E^2_{{\nu}_\tau(\bar{\nu}_{\tau})}}} N(Q^{2})~ dQ^2,
\end{equation}
with $a = \cos \theta_{c}~(\sin \theta_{c})$ for $\Delta S=0~(1)$ processes.

\subsection{Polarization observables of the final lepton}\label{pol}
 \begin{figure}
 \begin{center}  
        \includegraphics[height=5cm,width=10cm]{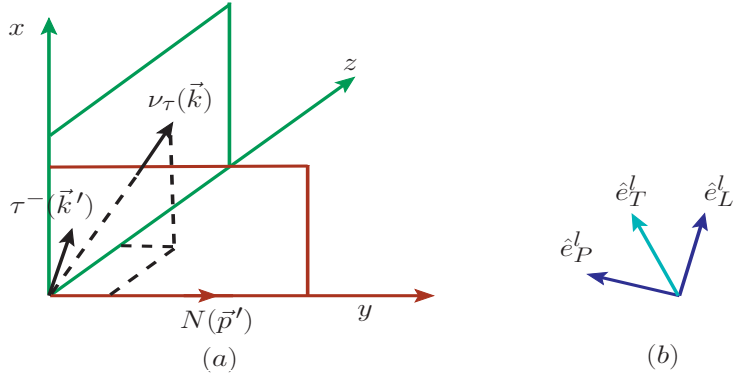}
  \caption{(a) Momentum and polarization directions of the final lepton produced in the reaction ${\nu}_{\tau} (k) + n (p) 
  \longrightarrow \tau^{-} (k^{\prime}) + p (p^{\prime})$. (b)~$\hat{e}_{L}^{l}$, 
  $\hat{e}_{P}^{l}$ and $\hat{e}_{T}^{l}$ represent the orthogonal unit vectors corresponding to the longitudinal, 
  perpendicular and transverse directions with respect to the momentum of the final lepton.}\label{T invariance}
   \end{center}
 \end{figure}
If one assumes the final lepton to be polarized, then the polarization 
4-vector~($\zeta^l$; $l=\tau$) is written as
\begin{eqnarray}\label{polar4}
\zeta^{l}=\frac{\mathrm{Tr}[\gamma^{\tau}\gamma_{5}~\rho_{f}(k^\prime)]}
{\mathrm{Tr}[\rho_{f}(k^\prime)]},
\end{eqnarray}
and the spin density matrix for the final lepton $\rho_f(k^\prime)$ is given by 
 \begin{equation}\label{polar1l}
 \rho_{f}(k^\prime)= {\cal J}^{\alpha \beta}  \text{ Tr}[\Lambda(k') \gamma_\alpha (1 \pm \gamma_5) \Lambda(k) 
 \tilde\gamma_ {\beta} (1 \pm \tilde\gamma_5)\Lambda(k')], 
\end{equation}
with $\tilde{\gamma}_{\alpha} =\gamma^0 \gamma^{\dagger}_{\alpha} \gamma^0$ and 
$\tilde{\gamma}_{5} =\gamma^0 \gamma^{\dagger}_{5} \gamma^0$. 

With $\rho_{f}(k^\prime)$ defined in Eq.~(\ref{polar1l}), the expression for $\zeta^{l}$ becomes
\begin{equation}\label{polar4l}
\zeta^{l}=\left( g^{\tau\sigma}-\frac{k'^{\tau}k'^{\sigma}}{m_\tau^2}\right)
\frac{  {\cal J}^{\alpha \beta}  \mathrm{Tr}
\left[\gamma_{\sigma}\gamma_{5}\Lambda(k') \gamma_\alpha (1 \pm \gamma_5) \Lambda(k) \tilde\gamma_ {\beta} (1 \pm 
\tilde\gamma_5) \right]}
{ {\cal J}^{\alpha \beta} \mathrm{Tr}\left[\Lambda(k') \gamma_\alpha (1 \pm \gamma_5) \Lambda(k) \tilde\gamma_ {\beta} 
(1 \pm \tilde\gamma_5) \right]},
\end{equation}
where $m_\tau$ is the mass of the $\tau$ lepton, $\Lambda(k) = \not{k}$, $\Lambda(k^{\prime}) = \not{k}+m_{\tau}$ and ${\cal J}^{\alpha \beta} $ is the hadronic tensor.

In the laboratory frame where the initial nucleon is at rest, the polarization 
vector $\vec{\zeta}$, is calculated to be a function of 3-momenta of incoming (anti)neutrino $({\vec{k}})$ and outgoing 
lepton $({\vec{k}}\,^{\prime})$, and is given as  
\begin{equation}\label{3poll}
 \vec{\zeta} =\left[{A^l(Q^2)\vec{ k}} + B^l(Q^2){\vec{k}}\,^{\prime} + C^l(Q^2) \vec{k} \times \vec{k}^{\prime}\right], 
\end{equation}
where the expressions of $A^l(Q^2)$, $B^l(Q^2)$ and $C^l(Q^2)$ are given in Ref.~\cite{Fatima:2018tzs}.

The polarization vector $\vec{\zeta}$, obtained from Eq.~(\ref{polar4l}), can be written as
\begin{equation}\label{3poll}
 \vec{\zeta}=\zeta_{L} \hat{e}_{L}^l+\zeta_{P} \hat{e}_{P}^l + \zeta_{T} \hat{e}_{T}^l, 
\end{equation}
where $\hat{e}_{P}^{l}$, $\hat{e}_{L}^{l}$ and $\hat{e}_{T}^{l}$ are the 
unit 
vectors corresponding to the perpendicular, longitudinal and transverse directions~(depicted in Fig.~\ref{T invariance}), 
which are given as
\begin{equation}\label{vectors}
\hat{e}_{L}^l=\frac{\vec{ k}^{\, \prime}}{|\vec{ k}^{\,\prime}|},\quad
\hat{e}_{P}^l=\hat{e}_{L}^l \times \hat{e}_T^l,\quad  
\hat{e}_T^l=\frac{\vec{ k}\times \vec{ k}^{\,\prime}}{|\vec{ k}\times \vec{ k}^{\,\prime}|}, 
\end{equation}
with $\zeta_{L,P,T}(Q^2)= \vec{\zeta} \cdot \hat{e}_{L,P,T}^l$.

The longitudinal $P_L^{l}(Q^2)$, perpendicular $P_P^{l}(Q^2)$ and 
transverse $P_T^{l}(Q^2)$ components of the polarization vector in
the rest frame of the outgoing lepton are obtained as:
\begin{eqnarray}
  P_L^{l} (Q^2) &=& \frac{m_\tau}{E_{k^{\prime}}} \zeta_{L}(Q^2) =  \frac{m_\tau}{E_{k^{\prime}}} \frac{A^l(Q^2) \vec{k}.
  \vec{k}^{\,\prime} + B^l (Q^2) 
  |\vec{k}^{\,\prime}|^2}{N(Q^2)~|\vec{k}^{\,\prime}|},\label{Pll}\\
 P_P^{l} (Q^2) &=& \zeta_{L} (Q^2) =  \frac{A^l(Q^2) [|\vec{k}|^2 |\vec{k}^{\,\prime}|^2 - (\vec{k}.\vec{k}^{\,\prime})^2]}
 {N(Q^2)~
 |\vec{k}^{\,\prime}| ~ |\vec{k}\times \vec{k}^{\,\prime}|},\label{Ppl}
  \end{eqnarray}
  \begin{eqnarray} 
  P_T^{l} (Q^2) &=& \zeta_{T} (Q^2) = \frac{C^l(Q^2) M [(\vec{k}.\vec{k}^{\,\prime})^2 - |\vec{k}|^2 |\vec{k}^{\,\prime}|^2]}
  {N(Q^2)~
  |\vec{k} \times \vec{k}^{\,\prime} |}, \label{Ptl}
\end{eqnarray}
where $\frac{m_\tau}{E_{k^{\prime}}}$ is the Lorentz boost factor, which appears due to the fact that we are measuring 
the polarization observables in the rest frame of the $\tau$ lepton. 
    \begin{figure}
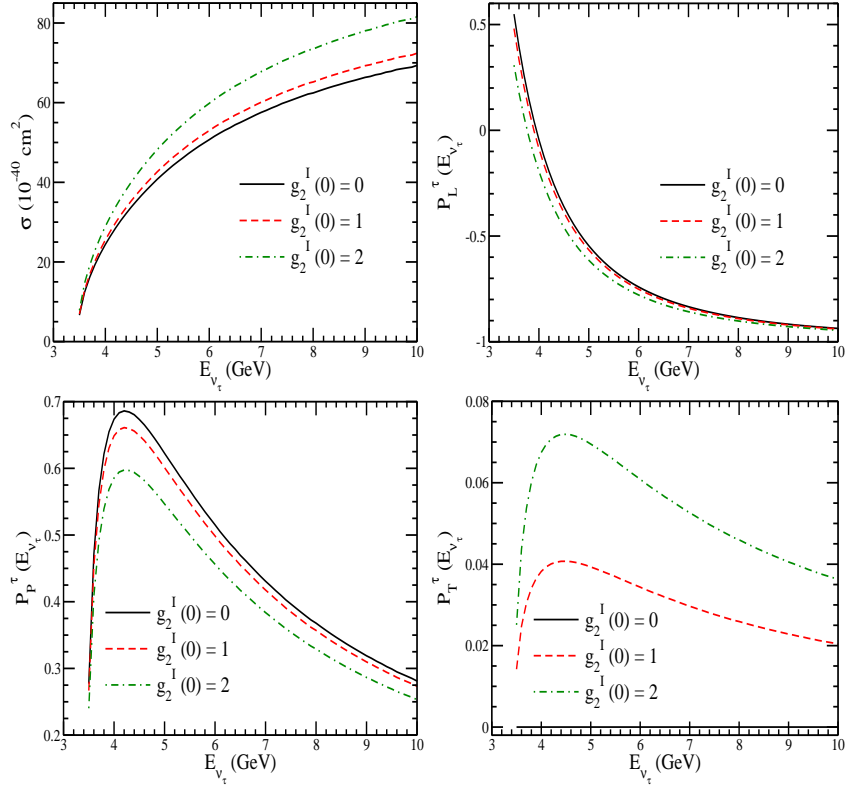

\centering
\includegraphics[height=5.2cm,width=5.5cm]{sigma_g2I_proton.eps}
\includegraphics[height=5.2cm,width=5.5cm]{Pl_g2I_proton.eps}
\includegraphics[height=5.2cm,width=5.5cm]{Pp_g2I_proton.eps}
\includegraphics[height=5.2cm,width=5.5cm]{Pt_g2I_proton.eps}
\caption{$\sigma$~(upper left panel), $P_{L}^{\tau} (E_{\nu_{\tau}})$~(upper right panel), $P_{P}^{\tau} 
(E_{\nu_{\tau}})$~(lower left panel) and $P_{T}^{\tau} (E_{\nu_{\tau}})$~(lower left panel) vs. $E_{{\nu}_{\tau}}$ for the 
process $\nu_{\tau} + n \rightarrow \tau^{-} + p$, using the different values of $g_{2}^{I} (0)$ {\it viz.} $g_{2}^{I} (0) 
= 0$~(solid line), 1~(dashed line) and $2$~(double-dotted-dashed line).}\label{sigma_g2I_nu}
\end{figure}
\section{SU(3) symmetry breaking}\label{SU}
The SU(3) symmetry breaking effects are incorporated following the works of Faessler {\it et 
al.}~\cite{Faessler:2008ix} and Schlumpf~\cite{Schlumpf}. In the following, we will discuss in brief the main features of 
these models.
\subsection{Model of Faessler {\it et al.}~\cite{Faessler:2008ix}}
Faessler {\it et al.} have studied the SU(3) symmetry breaking effects 
 on $f_{2} (Q^2)$ and $g_{1}(Q^2)$ form factors, using the constituent quark model. $f_{1}(Q^2)$ recieves no contribution 
 from the SU(3) symmetry breaking, at the leading 
 order because of the Ademollo-Gatto theorem. $g_{3}(Q^2)$ recieves the SU(3) breaking effects via $g_{1} 
 (Q^2)$. In this model, the form factors $f_{2} (Q^2)$ and $g_{1}(Q^2)$ are modified, at 
 $Q^2 =0$ in a similar manner {\it i.e.}, $f_{2}^{SU3} =g_{1}^{SU3} = {\cal F}$, which are given as:
  \begin{eqnarray}\label{SU3} 
{\cal F}^{p\Lambda}(0) &=& 
  -\sqrt{\frac{3}{2}} \left(F + \frac{D}{3} +\frac{1}{9} \left(H_{1} -2H_{2} - 3H_{3} -
   6H_{4}\right) \right),
   \end{eqnarray}
   \begin{eqnarray}
 {\cal F}^{n\Sigma^{-}}(0) &=& 
 D-F -\frac{1}{3}(H_{1} + H_{3}),
    \end{eqnarray}
    where $D$ and $F$ are the SU(3) symmetric  couplings while $H_{i}, i=1-4$ are the SU(3) symmetry breaking couplings, and 
    the value of these parameters for $f_{2} (Q^2)$ and $g_{1}(Q^2)$ form factors can be found in 
    Refs.~\cite{Faessler:2008ix, Fatima:2021ctt}.

\subsection{Model of Schlumpf~\cite{Schlumpf}}
    Schlumpf~\cite{Schlumpf} has studied SU(3) symmetry breaking effects on the vector $f_1 (Q^2)$ and axial 
 vector $g_1 (Q^2)$ form factors using relativistic quark 
 model. In this model, the modified $f_1 (Q^2)$ and $g_1 (Q^2)$ form factors are given by
 \begin{eqnarray}
  f_1^\prime(Q^2)=\alpha f_1(Q^2), \qquad
g_1^\prime(Q^2)=\beta g_1(Q^2),
 \end{eqnarray}
where $\alpha$=0.976 and 0.975;~$\beta$=1.072 and 1.056, respectively, for $p \longrightarrow 
  \Lambda$ and $n \longrightarrow \Sigma^-$ transitions~\cite{Schlumpf}.

\begin{figure}
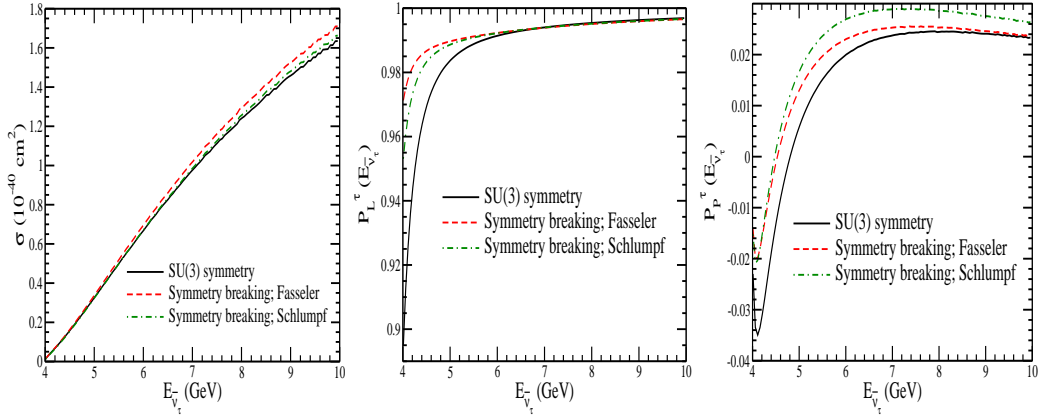

\centering
\includegraphics[height=5.5cm,width=4.5cm]{sigma_VFF_variation_lambda_SU3.eps}
\includegraphics[height=5.5cm,width=4.5cm]{Pl_VFF_variation_lambda_SU3.eps}
\includegraphics[height=5.5cm,width=4.5cm]{Pp_VFF_variation_lambda_SU3.eps}
\caption{$\sigma$~(left panel), $P_{L} (E_{\bar{\nu}_{\tau}})$~(middle panel), and $P_{P} (E_{\bar{\nu}_{\tau}})$~(right 
panel) vs $E_{\bar{\nu}_{\tau}}$ for $\bar{\nu}_{\tau} + p \rightarrow \tau^{+} + \Lambda$ process. The calculations have 
been performed using the SU(3) symmetry~(solid line), the SU(3) symmetry breaking effects parameterized by Faessler {\it et 
al.}~(dashed line) and by Schlumpf~(dashed-dotted line).}\label{sigma_VFF_lambda_SU3}
\end{figure}
\begin{figure}
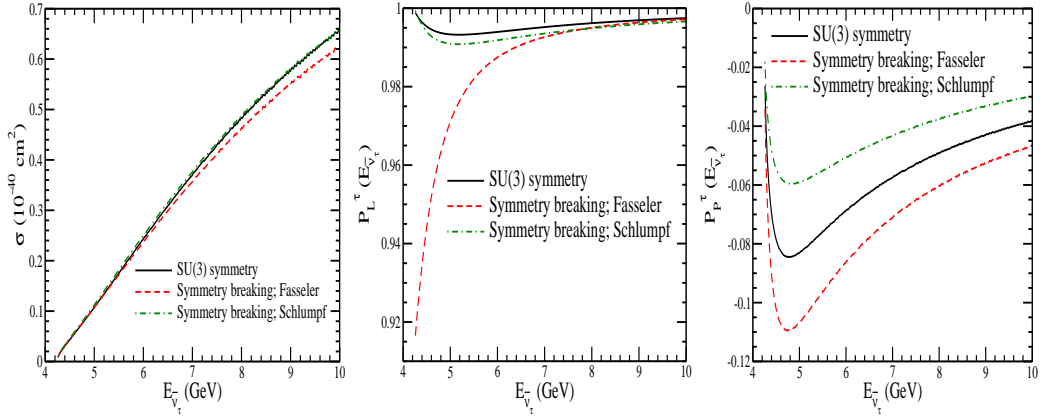

\centering
\includegraphics[height=5.5cm,width=4.5cm]{sigma_VFF_variation_SU3_sigma.eps}
\includegraphics[height=5.5cm,width=4.5cm]{Pl_VFF_variation_SU3_sigma.eps}
\includegraphics[height=5.5cm,width=4.5cm]{Pp_VFF_variation_SU3_sigma.eps}
\caption{$\sigma$~(left panel), $P_{L} (E_{\bar{\nu}_{\tau}})$~(middle panel), and $P_{P} (E_{\bar{\nu}_{\tau}})$~(right 
panel) vs $E_{\bar{\nu}_{\tau}}$ for $\bar{\nu}_{\tau} + n \rightarrow \tau^{+} + \Sigma^{-}$ process. The calculations have 
been performed using the SU(3) symmetry~(solid line), the SU(3) symmetry breaking effects parameterized by Faessler {\it et 
al.}~(dashed line) and by Schlumpf~(dashed-dotted line).}\label{sigma_VFF_sigma_SU3}
\end{figure}

\section{Results and Discussions}\label{results}
In Fig.~\ref{sigma_g2I_nu}, we have studied the effect of T violation on the total scattering cross section $\sigma 
(E_{{\nu}_{\tau}})$~(Eq.~(\ref{sig})), as well as on the polarization observables $P_{L}^{\tau} (E_{\nu_{\tau}})$, 
$P_{P}^{\tau} (E_{\nu_{\tau}})$ and $P_{T}^{\tau} (E_{\nu_{\tau}})$, for the process $\nu_{\tau} + n \rightarrow \tau^{-} + 
p$, by varying the value of $g_{2}^{I}(0)$ in the range $0-2$. It may be observed from the figure that the total cross section 
as well as the polarization observables~($P_{P}^{\tau} $ and $P_{T}^{\tau} $) are quite sensitive to the variation in the 
value of $g_{2}^{I} (0)$, while the effect of $g_{2}^{I} (0)$ variation on $P_{L}^{\tau} (E_{\nu_{\tau}})$ is small. 
The transverse component of polarization is non-zero when we 
take the T-violating effect~($g_{2}^{I} (0) >0$) into account. Therefore, the experimental measurement of the non-zero value 
of $P_{T}^{\tau}$, if any, would independently determine the degree of T violation in weak interactions. 

In Fig.~\ref{sigma_VFF_lambda_SU3}, we have presented the results for 
$\sigma(E_{\bar{\nu}_{\tau}})$, $P_L (E_{\bar{\nu}_{\tau}})$ and $P_P (E_{\bar{\nu}_{\tau}})$ vs 
$E_{\bar{\nu}_{\tau}}$ for the charged current process $\bar{\nu}_{\tau} + p \longrightarrow \tau^+ + \Lambda$ with SU(3) 
symmetry as well as when the SU(3) symmetry breaking effects are taken into account. We find that the effect of SU(3) 
symmetry breaking on 
$P_P (E_{\bar{\nu}_{\tau}})$ is quite significant, while on $P_L (E_{\bar{\nu}_{\tau}})$ the effect is small and there is 
hardly any effect of SU(3) symmetry breaking on $\sigma(E_{\bar{\nu}_{\tau}})$.

In Fig.~\ref{sigma_VFF_sigma_SU3}, we have presented the results for $\sigma(E_{\bar{\nu}_{\tau}})$, $P_L 
(E_{\bar{\nu}_{\tau}})$ and $P_P (E_{\bar{\nu}_{\tau}})$ vs $E_{\bar{\nu}_{\tau}}$ for the reaction $\bar{\nu}_{\tau} + n 
\longrightarrow \tau^+ + \Sigma^-$ with SU(3) symmetry as well as when the SU(3) symmetry breaking effects are taken into 
account. We find that 
in $\Sigma$ production, there is a large variation in $P_L (E_{\bar{\nu}_{\tau}})$ and $P_P (E_{\bar{\nu}_{\tau}})$ due to 
the SU(3) symmetry breaking while the effect on $\sigma(E_{\bar{\nu}_{\tau}})$ is small like the one observed in $\Lambda$ 
production. In the case of $P_L (E_{\bar{\nu}_{\tau}})$,  the variation is more when Faessler {\it et 
al.}~\cite{Faessler:2008ix} prescription is used and the nature of dependence on SU(3) breaking effect using Faessler {\it et 
al.}~\cite{Faessler:2008ix} and Schlumpf~\cite{Schlumpf}, prescriptions are different.

\section{Conclusion}\label{conclusion}

 The effect of T violation is 
  appreciable for $g_2^I(0)\ge 1$ in the case of $\sigma $ as well as the polarization observables both for the neutrino 
  as well as antineutrino induced processes. 
    There is a significant variation in $P_{L}$ and $P_{P}$, on the different parameterizations of the SU(3) 
    symmetry breaking, while the effect of SU(3) symmetry 
  breaking is not much in the case of total scattering cross sections.
   These results are more prominent in the case of 
     $\Delta S=1$ processes.



%

\end{document}